# An Anonymous On-Street Parking Authentication Scheme via Zero-Knowledge Set Membership Proof

Jerry Chien Lin Ho
*Department of Computer Science and Information Engineering*
*Tamkang University*
Tamsui, Taipei 251, Taiwan
a29988122@gmail.com

Chi-Yi Lin*
*Department of Computer Science and Information Engineering*
*Tamkang University*
Tamsui, Taipei 251, Taiwan
chiyilin@mail.tku.edu.tw

*Abstract*—The amount of information generated grows as more and more sensor and IoT devices are deployed in smart cities. It is of utmost importance for us to consider the privacy data leakage and compromised identity from both outside adversaries and inside abuse of data access privilege. The security assumption of the system should not solely rely on the fact that permission and access control were being implemented correctly. Quite the contrary, a system can be designed in a way that user's identity data and usage traces are not leaked even if the system had been compromised. Based upon our previous on-street parking system utilizing Bluetooth Low Energy (BLE) beacons, we applied a cryptographic primitive called zero-knowledge proof to our authentication system. A commitment scheme and Merkle tree is combined in the setup to achieve zero-knowledge set membership proof. Doing so, the user is anonymous to the server between authentication sessions, while the server's still able to verify the legitimacy of such user. The on-street parking system is therefore immune to privacy data leakage, as for now one cannot mass-query and profile certain user's traces within the system.

*Keywords—smart parking, zero-knowledge set membership proof, bluetooth low energy beacon*

## I. Introduction

There's an absurd amount of information being generated whenever a new communication happened between two nodes, especially in megacities where its citizens are tightly knitted. For example, mass volume of vehicle traffic data is gathered and relayed in real-time via smartphones and roadside sensors, in order to make dynamic traffic control and on-street parking allocation possible [1], [2].

However, given the overwhelming complexity of possible interactions between nodes, security and privacy protection of such systems are often overlooked or even omitted in the name of engineering practicality. It is also difficult for researchers to build a believable simulation environment to gather data and test systems before actual deployment of said system, therefore causing unforeseeable consequences [3], [4].

As a result, there were a few examples of compromised systems, rendering its user being tracked, vehicle and roadside units (RSUs) firmware being overwritten, or even vehicle's direction being manipulated via forged navigations [5], [6]. Autonomous vehicle fleets are no exception to this kind of issues [7]. To put things to an extreme, terrorists can fingerprint the target vehicle's network signal [8], [9], and trigger the bomb as the target vehicle approaches [6]. Even in the less severe cases, Wang et al. [10] attacked Waze to expose and pinpoint certain user, successfully tracked their location. As a result, new incidents of privacy data leaks happen day by day, harming citizens who are living in modern smart cities.

Apart from implementation oversight, it's also getting more and more common that the cause of privacy leak originates from an inside job rather than an outside attack [11]. Whether the leak is intentional or not, the misuse and abuse of data access privilege is a worrying trend in both private and public sectors [12]. The invasive personal data gathering is harmful to the privacy of its citizens, especially when the recorded data and the scale of such were not carefully examined [1], [13]. To make things worse, mass deployment of sensors and cameras is making mass surveillance an unfortunate reality – which means – the cost of profiling someone has never been so cheap, to adversaries with malicious intents.

Since the abovementioned, once far-fetched sounding scenario now becomes a reality, we should really start designing security systems with privacy-preserving features in mind. For example, a *zero-knowledge based authentication system* provided novel security features where its users do not expose identity nor usage traces when accessing the system, while the system's able to confirm the permission and the validness to access for such user.

Sun et al. [7] defined five major categories of IoV attacks:

- Attacks on authentication.
- Availability attack
- Secrecy attack
- Routing attack
- Data authenticity attack

From our previous work of *on-street smart parking* [14] shown in Figure 1, the system is vulnerable and prone to authentication attack and replay attack, as the identity string *major* and *minor* of BLE beacons are broadcasted publicly in *plaintext*. This renders eavesdropping and forging of vehicle's identity possible to adversaries [15], which also makes the accountability of the system in question. Although the use of BLE beacon is reasonable on the aspect of cost and performance in parking space detection, it is not suitable in the authentication scheme due to incompatible security properties [16], [17].

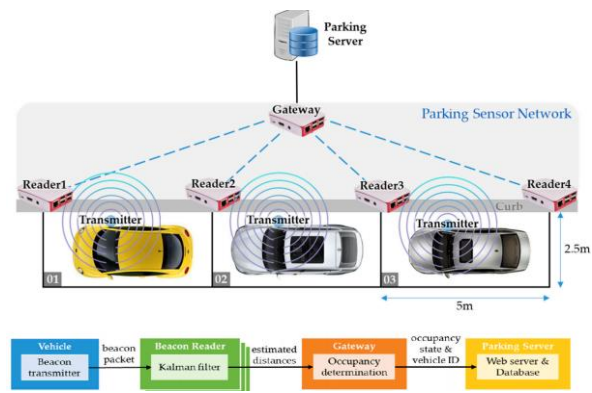

Fig. 1. The system architecture in our previous work.

\* Corresponding author

In order to improve the security of our system, we applied Gabizon's *zk-SNARK* variant [18] combining with a commitment scheme [19] and a *Binary Hash Tree* (a.k.a. *Merkle tree*) [20] onto our system to authenticate users. The validity of arbitrary user can therefore be verified without compromising his / her identity, while the two-way confidentiality between vehicles and *RSUs* are improved as a result, which also eliminates the possibility of identity forgery in our previous work.

Doing so, the privacy of users can be preserved in a way that even accidental data leak or data breach cannot compromise the parking data and traces of users, as the system never knows who accessed the system at a certain point – the system only knows that the user's eligible accessing.

Compared to existing solutions [21]–[34], our system excels in the following aspects:

- The performance of proof construction is the same, in some cases, better, compared to the current zero-knowledge or signature-based solutions, which makes it relatively hassle-free to be implemented on an *OBU*.
- The protocol is non-interactive once initiated – which means it's more suitable to the unstable outdoor Bluetooth and cellular network environment.
- The solution provided novel security properties like *zero-knowledge proof of membership*.
- The small packet size generated makes it suitable in an ad-hoc Bluetooth environment.
- Most importantly, the proposed system prevents insider threats from identity and privacy data leakage, while other systems do not take this aspect into consideration.

The rest of the paper is organized as follows. In Section II, we'll be listing existing privacy and security solutions in the context of smart city applications. The pros and cons of such solutions are listed and then compared to our zk-SNARK based solution. In Section III, the cryptography primitives used in our paper will be introduced. In Section IV, we'll describe our solution in detail. In Section V, the security and performance aspects of our system as well as other existing solutions will be examined closely. Section VI gives some discussions as well as future directions. Finally, Section VII concludes our work.

## II. LITERATURE REVIEW

In this paper, we aimed to improve the security aspects of our previous research of the low-cost on-street parking system. Replacing the BLE authentication scheme with zk-SNARK, we achieved the properties of computationally hiding and perfect binding of user's identity commitment. The following paragraph of this section will be reviewing the topics in the following order: the peculiar usages of *zero-knowledge proof*, PKI-based data anonymization, *zero-knowledge proof* for authentication, and the contributions of our construction.

### A. Off-Label Uses of Zero-Knowledge Proof

One could falsely believe that *zero-knowledge proof* can only be used for privacy data protection and identity anonymization. In fact, just as all the cryptographic primitives have their "off-label use", *zero-knowledge proof* can also be used for data aggregation, data availability, or validity proof in the domain of *secure multi-party computation(sMPC)*.

Kursawe et al. proposed a privacy-friendly data aggregation scheme for smart-grid [31]. It hides the individual meter's usage statics while still preserving the ability to detect pipe leakage or bill fraud. It is also compatible with statistical analysis for energy grid optimization. The use of *zero-knowledge proof* in this paper is to prove the committed random value and the reported meter reading are genuine.

In the landscape of blockchain – a promising system for *sMPC*, the balance between data availability, data validity, and network throughput is always an issue. To increase the network throughput, the most intuitive solution is to offload some *transactions* (state transition functions) onto another layer of the network, rather than processing all the computations on-chain. In this case, the ability to check the validity and the integrity is crucial to the robustness of the system. ZK rollup and optimistic rollup [35] aimed to solve this issue by providing a *zero-knowledge proof* for batch processing of *transactions* to ensure the data validity of the result, either opportunistically or deterministically, while making each of the transaction data available on-chain. Benefiting from *zero-knowledge proof*, ZK rollup and optimistic rollup solved the data availability issue in other solutions such as plasma [36] or state channel [37]. The use of *zero-knowledge proof* here also works as sort of a compression, which saves storage compared to the space raw transaction data occupied.

### B. Data Anonymization

Guo et al. [32] proposed a way to securely collect a mass volume of data in the IoV network. It utilizes a traditional PKI-based approach, while the identity anonymization and privacy of users are not considered on the server-side. Sun et al. [33] came out with a blockchain-based solution that enables the VSN data to be accessed by authorized third party vehicle data users, while preventing the data to be tampered with. However, it does not prevent the misuse of such data from authorized parties i.e. an insider. Ou et al. [34] proposed a Zcash based data anonymization scheme for vehicular data transactions and trading. Although it achieved a reliable anonymous data transaction, the traded data itself could still contain some sensible tracking information from its originated vehicle and its owner, which is harmful to the privacy of citizens.

### C. Authentication

Liu et al. [21] described a key agreement scheme to achieve mutual vehicle authentication in the V2V network. The proposed system protected privacy between vehicles, anonymized their identity. However, the described protocol is not suitable for our parking system, as the protocol complexity is too high. Jo et al. [22] proposed an anonymized ECU-to-ECU authentication system to prevent masquerade attack and identity forgery. The paper constructed a CAN-based authentication scheme to prevent illegal access. The system is not suitable to our parking scenario, as the main focus of our construction is on the unlinkability for a certain user between parking sessions.

Gope et al. [23] proposed a privacy-preserving key agreement scheme between drones and service providers. The system achieved the same level of privacy and security as our system does. However, it utilized a traditional key-agreement approach, rather than our *zero-knowledge proof* approach.

Multiple rounds of challenge and response interactions are also needed between drones and service providers, and the computational cost is slightly higher than our approach. Zhu et al. [29] also proposed an anonymous smart-parking and payment scheme. The paper deployed a traditional short randomizable signature to provide anonymity and conditional privacy. The system aimed to achieve privacy protection of vehicles from third-party parking service providers, which is also the main focus of our paper. However, they designed a "fall-back" mechanism in their system. The identity of the vehicle can be revealed whenever a dispute happens – that gives malicious insiders or outside adversaries chances to compromise user's privacy, which should never happen in the first place.

For *zero-knowledge proof* based approach, Haddad et al. [24] protected user's identity and location privacy in mobile LTE network via the famous Schnorr protocol of *zero-knowledge proof*. Their setup is out-of-date, thus vulnerable in a sense. It's also an interactive process rather than non-interactive process. Soewito et al. [25] proposed a zk-PoK (*zero-knowledge Proof of Knowledge*) scheme for anonymous login on websites. Their protocol was based on a 1998 setup, thus not an ideal choice for authentication on the aspect of both security and efficiency. Han et al. [26] applied *Feige-Fiat-Shamir zero-knowledge proof* for the authentication between trusted authority, user's mobile phone, and vehicle. This is a challenge-response based protocol, which consists of multiple interactive rounds. The setup also heavily relied on trusted authority, which amplifies privacy issues.

### D. Similar Works

Among all the smart city authentication schemes, the work from Gabay et al. [27], and Huang et al. [30] are constructed similarly to our system, both on the used cryptographic primitives and the intended application scenario. The performance of the abovementioned works will be compared later in the paper.

Gabay et al. [27] proposed an interesting design for EV charging and payment system, based on Zcash and blockchain [38]. It protected EVs from revealing their identity to arbitrary charging stations, while still preserving the ability for payment and identity authentication. The setup relies on a private or public blockchain network, which is not the case in our system. Although the system achieved mutual authentication between charging stations and EVs, the system is, in a sense, overengineered for our on-street parking system. A blockchain-based approach assures the data integrity on chain, there's no single party who can modify the data without proper permission. However, in our on-street parking scenario, users do not need to be worried about other user's identity – he/she only cares about their own identity. In this case, authorities should be able to modify the identity database freely, as this is the intended design rationale.

Walshe et al. [28] proposed a novel IoT authentication scheme for IoT devices, in the hope of preventing MitM attack on IoT gateways. It utilizes the same Merkle tree opening traversal techniques to our system. The method used in their system is mainly used to prevent MitM attack, rather than preventing privacy data leakage. As a result, the protocol used a whole Merkle tree to prove the identity of IoT nodes during the interactive authentication process. This compromised part of the privacy of the IoT devices, and is not suitable for our on-street parking scheme.

Huang et al. [30] applied an interactive Fiat-Shamir heuristic zk-PoK, based on a random oracle model. The paper proposed an automated valet parking scheme, prevented double-reservation attack, and achieved vehicle anonymity to the third-party parking provider in the process. The system assumed autonomous vehicles and a designated parking lot management system, which is far more limited than our system. The vehicle is authenticated via a token-based approach, which expired after 24 hours, which is different from our partially reusable zk-SNARK proof.

### E. Review The Whole Picture

In the domain of IoT and IoV, various authentication scheme with different security properties and system complexity has been applied [39]. The majority of them were occupied with traditional PKI-based approach, while some utilized end-to-end authentication and encryption[21]–[23].

In a sense, only *battle-tested* cryptographic primitives were seen and widely used in IoT and IoV domain, while novel state-of-the-art cryptographic primitives such as sMPC, zero-knowledge proof, homomorphic encryption, or blockchain-based approach which focuses on the privacy protection and anonymity were infrequently seen. Among those which deployed zero-knowledge based authentication schemes [24]–[34], they were either based on *interactive* zero-knowledge schemes, out-of-date zk constructions, or even vulnerable – such as malleable – zk variants.

Therefore, we presented a novel non-interactive zero-knowledge [18] based authentication scheme utilizing Bluetooth connection for the following reasons:

- The zk-SNARK variant is non-interactive, thus more robust and reliable in the IoT and IoV environment.

- The required computational power and proof size is well-balanced in a sense suitable to our Bluetooth-based roadside parking scheme.

- The security and communication requirement in our system is discrepant from other systems with more complex interactions. Our redesigned zk-SNARK scheme can be deployed relatively with ease.

- The vehicle is anonymized without any "strong assumption" – the proof system is computational sounding, which means that a dishonest prover cannot construct such a proof to persuade a verifier to believe a false statement i.e., the adversary is polynomially bounded.

## III. PRELIMINARIES

The core concept to *zero-knowledge* and *proving system* lies in the area of computational complexity theory. How does it achieve all the abovementioned properties, and how do we make use of them in our authentication system?

To answer all the questions, as to what a *zero-knowledge proof* really is, we must explore its origin dated back to 1989 – an *interactive proof system* [40]. The following concepts in the domain of computational complexity will be introduced in the following paragraphs: NP language, Boolean satisfiability problem, and proving systems.

### A. NP Language

The solution to a given NP problem should be efficiently verifiable by a deterministic Turing machine in polynomial time. Arora and Barak defined NP class and NP Language in their book *Computational Complexity: A Modern Approach* [41] as follows:

**Definition 3.1** (NP Class).

A language $\mathcal{L} \subseteq \{0,1\}^*$ is in NP, if there exists a polynomial p : $\mathbb{N} \to \mathbb{N}$ and a polynomial-time Turing machine $M$ (called the verifier for $\mathcal{L}$), such that for every $x \in \{0,1\}^*$,

$$x \in \mathcal{L} \Leftrightarrow \exists\, u \in \{0,1\}^{p(|x|)} \; s.t. \, M(x,u) = 1$$

That is, if $x \in \mathcal{L}$ and $u \in \{0,1\}^{p(|x|)}$ satisfy $M(x,u) = 1$, we can call $u$ a certificate (or witness) for $x$, with respect to the language $\mathcal{L}$ and the machine $M$. Note the term witness, as it will be later used in the zk-SNARK setup.

One can find out that this scenario is close to the zk-SNARK setup – the prover holds extra information witness $u$ in order to satisfy the statement $x \in \mathcal{L}$ to be true. However, the process still requires $u$ to be seen during verification, hence not a *zero-knowledge* protocol.

*B. Boolean satisfiability problem (SAT), Circuit satisfiability problem (CircuitSAT), Quadratic Arithmetic Programs, and Arithmetic circuits*

Apparently, we need to construct our NP language $\mathcal{L}$ in a way that the argument of the "circuit" can be carefully crafted to fit our needs.

Consider the most basic form of such problem called *Boolean satisfiability problem (SAT)*. Consisting a set of propositional variables $x$, and the boolean connectives ¬, ∧, and ∨, the resulting circuit can be called *boolean formula*, denoted as $F$.

**Definition 3.2** (SAT problem).

$$SAT(F(x)) = 1 \Leftrightarrow F \text{ is satisfied given witness } u = x$$

SAT is NP-complete [42], [43], and that means there exists a reduction function $f$ where the following relationship holds true for any NP Language $\mathcal{L}$ :

**Definition 3.3** (SAT reduction).

$$\mathcal{L}(x) = SAT(f(x))$$

The insight to this reduction process is that, we can choose *any* NP Language that suits our need in *zero-knowledge proving protocols* – reduce it and still be able to verify it in polynomial time.

Such construction was proposed by Gennaro et al. [44]. In the form of polynomials, the so-called *Quadratic Span Programs* and *Quadratic Arithmetic Programs* were based on CircuitSAT and arithmetic circuit, respectively.

Instead of taking variables, CircuitSAT is taking circuit as inputs. CircuitSAT outputs true if and only if the circuit $C$ is *satisfiable* i.e., there exists at least one input value that $C(x) = 1$. Nitulescu defined a circuit satisfaction problem as follows [45]:

**Definition 3.4** (Circuit Satisfaction CircuitSAT).

The circuit satisfaction problem of a circuit $C$:

$$I_u \times I_w \to \{0,1\}$$

is defined by the relation:

$$R\_C = \{(u,w) \in \{u \in I\_u \,:\, \exists\, w \in I\_w \,,\, C(u,w) = 1\}$$

CircuitSAT is the foundation of zk-SNARK, as it allows arbitrary high-level algorithms and arguments to be constructed into satisfiable and verifiable circuits.

Effectively working as the input to CircuitSAT, arithmetic circuit is a natural way to represent polynomials. Raz defined an arithmetic circuit [46] as follows:

**Definition 3.5** (Arithmetic circuit).

Let $\mathbb{F}$ be a field, and let $\{x_1, \ldots, x_n\}$ be a set of input variables. An arithmetic circuit is a directed acyclic graph, where each root vertex is labeled by either a variable $x_i$ or a field element $\alpha \in C$, and each non-root vertex is labeled by either a + or × sign. Every edge in the graph is labeled with an arbitrary field element. A node of out-degree 0 is called an output-gate of the circuit.

The polynomial form of arguments is an efficient way to "compact" the original statements, and the conversion process reduced the needed constraints/gates i.e., conditional expressions to be verified. This process is called *arithmetization* [47]. Gennaro et al. [44] introduced *Quadratic Span Programs* and *Quadratic Arithmetic Programs* for this conversion process, which is then used in Groth's *CRS* construction [48]. In our proposed system, we used the construction from Gabizon et al. [18], which provided better efficiency and maintainability of the circuit than the R1CS constraint and the QAP system used in Groth's previous work [48].

*C. Proving System*

Convert the provable NP statement into polynomial expressions and arithmetic circuit is one thing; Interact with the circuit and convince the verifier with zero-knowledge, completeness, soundness, and succinctness is another.

Arora et al. [49] proposed a probabilistic checkable proof NP class. In PCP, the verifier is able to access and verify the proof string in a random access manner, only reads a constant number of bits from the proof. Arora et al. [49] defined the PCP as follows:

**Definition 3.6** (PCP).

For functions $r, q: \mathcal{L}^+ \to \mathcal{L}^+$, a probabilistic polynomial-time verifier $V$ is $(r(n), q(n)) - restricted$ if, for every input of size $n$, it uses at most $r(n)$ random bits and examines at most $q(n)$ bits in the membership proof while checking it. A language $\mathcal{L} \in PCP(r(n), q(n))$ if there exists a $(r(n), q(n)) - restricted$ polynomial-time verifier that, for every input $x$, behaves as follows:

- if $x \in \mathcal{L}$, then there exists a membership proof $\pi$ such that $V$ accepts $(x, \pi)$ with probability 1 (i.e., for every choice of its random bits);

- if $x \notin \mathcal{L}$, then for any membership proof $\pi, V$ accepts $\pi$ with probability at most $\frac{1}{2}$ .

In order to have a NIZK proving system rather than an interactive PCP-constructed proving system, Gabizon19 proposed a construction, which based on the Fiat-Shamir heuristics, with polynomials, commitments to polynomials

(Kate commitments specifically), wire and gate permutation challenges, and bilinear parings on eliptic curve such as BN254, to achieve the property of homomorphic hiding in order to accelerate the verification effort while maintaining the zero-knowledge property. According to Gabizon [18], the property of zero-knowledge depends on the amount of information in the trusted party $I$ leaked into the final compacted protocol. In order to achieve this property, Gabizon added random multiples of $Z_H$ into the witness based-polynomials, where $H$ is a multiplicative subgroup containing the nth roots of unity in $\mathbb{F}_P$, the primitive nth root of unity, and a generator $\omega$ of $H$, as in the form: $H = \{1, \omega, ..., \omega^{n-1}\}$. This makes a verifier unable to obtain extra information in the verification process when opening the commitments and evaluating polynomials.

## IV. PROPOSED SYSTEM MODEL

### A. Overview

We proposed an on-street parking authentication system focusing on the protection of user and vehicle privacy. Figure 2 shows the system architecture. Applying a succinct non-interactive argument of knowledge zero-knowledge proof of user validity [38], [50] during the authentication process of parking, we can effectively prevent a mass surveillance Orwellian government from happening, as one cannot gather and accumulate the access data of certain vehicle and its user explicitly on the server-side.

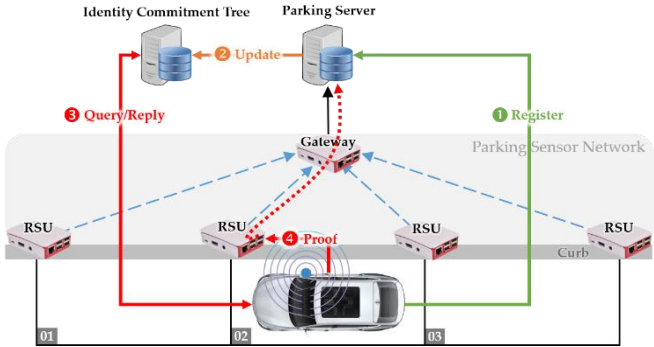

Fig. 2. System Architecture.

zk-SNARK is a variant of *zero-knowledge proof*, proposed by Groth [48]. It excels its predecessors on the succinctness. It has relatively small proof sizes and minimal verifying effort for the verifier. It is also of a non-interactive (NIZK) construction.

Groth defined the following properties in zk-SNARK [48]:

- Completeness: Given a statement and a witness, the prover can convince the verifier.
- Soundness: A malicious prover cannot convince the verifier of a false statement.
- Zero-knowledge: The proof does not reveal anything but the truth of the statement, in particular it does not reveal the prover's witness.

For the definition of interactivity, Blum et al. derived a non-interactive set-up from a *common reference string* model [51]:

- Non-interactive: A prover does not need to communicate with the verifier beforehand when constructing his / her proof.
- Monodirectional: Utilizing the common reference string (CRS) model, the proof can be sent one-way from the prover to the verifier.

Groth et al. [52] and Maller et al. [53] later updated the *common reference string* model into a universally and continually updatable *structured reference string* (SRS) model. This makes the predefined circuit setup more convenient to be deployed and maintained.

For our system, we applied Gabizon's *zk-SNARK* variant [18] with the SRS setup, which proved to be a good trade-off between proof size, proof time, and trusted setup maintainability.

Table I lists the notations used in the system. The system has two phases. In the registration phase, users must register their vehicle data and national ID, along with the hash of their public key as an identity commitment to the authority server. $H(PU_U)$ is then stored in an identity commitment Merkle tree.

**Table I: Notations**

| Notation | Definition |
|---|---|
| UID | User's National ID, and the corresponding license data |
| $PU_U$ | User's public key |
| $PR_U$ | User's private key |
| $H(\ )$ | Poseidon hash function |
| Vehicle | User's vehicle, with bluetooth connection capabilities |
| Server | Parking server, store data and authenticate user validity |
| RSU | Road side units, maintain connections between server and vehicles |
| C | An arithmetic proving circuit |
| pk, vk | Proving key and verifying key of the circuit |
| $\pi$ | ZK proof generated by user |
| nu | Nullifier, a rotating nonce broadcasting by server |
| mp | Merkle path from user commitment leaf to root |
| rh | Root hash of the Merkle tree |
| $l_{id}$ | The leaf index where user stores identity commitment |

In the authentication phase, users would query the server to look up the values needed to construct the proof, which includes the current position of their identity commitment leaf $l_{id}$ in the identity Merkle tree, the path to the root $mp$, and the Merkle root of the tree $rh$. Users also retrieve a rotating random value from server $nu$ called *nullifier*, in order to prevent the double singling of such zero-knowledge proof. Users can then construct their proof using *plonk converted* [18] arithmetic circuit and private inputs consisting of $(PR_U, mp, rh, nu, pk, vk)$. The proof $\pi$ is then sent to the $RSU$ in order to prove that the users are eligible and authorized to park.

### B. Registration Phase

Users are required to register their National ID, vehicle registration information, vehicle plate number, and other data required as per the authority requested. Users should send the aforementioned information alone with their hashed public key $H(PU_U)$ as *identity commitment*, in order to complete the registration process. *Identity commitment* is then stored on $Server$ in the form of incremental Merkle tree. The leaf index $l_{id}$ and the current Merkle path $mp$ is then transferred back to the vehicle for proof generation in the following phase. Algorithm 1 shows the whole registration process.

**Algorithm 1:** Registration Phase.
1:   Server.register() ← Commitment(H(PU$_U$), UID)
2:   Server.store()
3:     MT = MerkleTree().create
4:     MT.InsertLeaf(Commitment(H(PU$_U$), UID))
5:     Roothash.Update(MT)
6:   return
7:   Vehicle ← Msg(leaf, mp, rh)

Although one's personal information is stored in the registration phase, it is unable to link one's authorization request to the identity registered in the database, as zero-knowledge proof allows the confirmation of user validity without revealing user's identity. The design rationale is to intentionally raise the hurdle to link personal data with vehicle footprint, whereas keeping the aspect of accountability in check. One must physically and visually examine the vehicle's plate number in order to know which user is currently occupying the parking space, thus making mass accumulation of certain user's vehicle traffic data infeasible.

As a one-time setup, the *Server* must construct the arithmetic circuit $C$ [54], generate $prove()$ and $verify()$ function, and $vk, pk$ pair used in *zero-knowledge proving system*.

**Definition 4.1** (proving system setup).

$$prove(), verify(), (pk, vk) \leftarrow G(\lambda, C)$$

*C. Authentication Phase*

In the authentication phase, users must query the current leaf index $l_{id}$ of their *identity commitment*, the corresponding $mp$, and the $rh$ of the *incremental Merkle tree* from arbitrary data sources. In order to prevent Server's attempts on revealing users' identity via time-based tagging [55], it is suggested for users to retrieve $mp$ beforehand, or even maintain and sync their own copy of the *incremental Merkle tree*.

**Algorithm 2:** Proof Construction.
1:   *Private Input:*
2:     w ← (l$_{id}$, mp, H(PR$_U$))
3:   *Public Input:*
4:     x ← (rh, nu)
5:   *procedure:*
6:     π ← Prove(pk, x, w)
7:     Server.auth() ← Msg(π)

As seen in Algorithm 2, the retrieved data is then used to construct a proof $\pi$, ready to be sent to the Server, relaying via *RSU* by means of Bluetooth connection.

The proof is constructed using the latest $nu$, which is rotated server-side, thus the proof cannot be reused or forged by adversaries. The proof itself cannot be associated with certain user, as they are *computationally indistinguishable* from each other. That is, neither adversaries can fingerprint certain user nor steal their identity.

The constructed arithmetic circuit proves the following three things:

- $H(PU_U)$ is the corresponding public key to $H(PR_U)$, and the prover has the knowledge to $PR_U$.

- The $nu$ is up-to-date, and the proof using such $nu$ has never been seen before server-side. As long as the *Server* sees a proof with an out-of-date $nu$ or a used $nu$, the authentication request is rejected.

- The prover has the knowledge to the *identity commitment* of certain $l_{id}$ in the *identity commitment tree*, as the prover can build a proof which concludes the correct $rh$ and $mp$, from which the user's leaf resides.

As the server receives the proof, the $verify()$ function as seen in Algorithm 3 is called in order to check the validity of user's authentication request.

Once the proof $\pi$ is verified to be *true*, the Server returns the result to the origin *RSU* near user vehicle, authorizes user's parking request.

**Algorithm 3:** Proof Verification.
1:   *Public Input:*
2:     x ← (rh, nu)
3:   ZK.Proof(π)
4:     result ← Verify(vk, x, π)
5:     if result == TRUE
6:       return TRUE
7:     else
8:       return FALSE
9:   return

The security for the set membership proof relies on a collision resistant hashing (CRH) algorithm. For the form $H(x) = h$, the pre-image $x$ is the witness of the CRH function $H()$. Recall the definition 3.2, we can reach the conclusion that hashing functions are a kind of satisfiability problem [56], as such a witness $x$ exists for the hashing function that can be verified in polynomial time. In Gabizon's construction, the hashing function is represented as an arithmetic circuit over a large prime field. The circuit used in our proposed system implemented a CRH function called Poseidon hash [57], where its efficiency in the arithmetic circuit is superior to other CRH functions such as Pedersen hash, MiMC hash, or SHA-256, on the circuit constraint count and the running time taken.

To further explore the role of Merkle tree in set membership proof, Dahlberg et al. [58] have given a clear explanation: "... for the case of a single SMT with a fixed hash function, no special encoding is necessary to distinguish between nodes, and that the security of an audit path reduces to the collision resistance of the underlying hash function." In other words, the calculation of authentication paths is the process of *hash-chaining*. That is, once the identity commitment is committed to the leaf and the root hash is derived, one cannot efficiently persuade and generate proof of the authentication path for different leaf values (commitments), given the hashing function is a CRH function.

V. PERFORMANCE ANALYSIS

In the following paragraph, we'll list the theoretical performance of different proof systems, such as Groth16's snark [48], Gabizon19's plonk [18], and Huang18's zk-PoK

[30]. For the actual performance, we'll list Gabay20's work [27], Huang18's work [30], and our work for comparison.

### A. Proof System Performances

The performance of a proving system depends on the multiplication and exponentiation operations done on different groups. The notation $\mathbb{G}$ stands for groups, $\ell$ stands for statement lengths, $m$ stands for the number of wires, $n$ stands for the number of multiplication gates, and $a$ stands for the number of addition gates. Table II shows the theoretical performance of the underlying proving system.

**Table II: Proof System Performance Comparison**

|  | Proving Performance | Proof Size |
|---|---|---|
| Groth16 [48] | $3n + m - \ell \; \mathbb{G}_1 \; exp, n\mathbb{G}_2 \; exp$ | $3n + m\mathbb{G}_1$ |
| Huang18 [30] | $3\mathbb{G}_T \; exp + 4\mathbb{G}_T \; exp + 8\mathbb{P}_p$ | N/A *1 |
| Gabizon20 [18] | $9n + 9a \; \mathbb{G}_1 \; exp,$ $\approx 54(n+a) \log(n+a) \; \mathbb{F} \; mul$ | $n + a\mathbb{G}_1, 1\mathbb{G}_2$ |

[1] Proportional to reserved user count.

On average, Groth16 is 1.1 ~ 2 times faster than Gabizon20 on prover time. For example, A pre-image proof circuit of SHA-256 took 2.1s for Gabizon20, and 1.4s for Groth16 [59]. However, Gabizon20 is far more superior for its universal updatable *structured referenced string*, as this prevents the hassle of doing a trusted setup all over again when updating the circuit design. Also, by applying "circuit/snark friendly cryptographic primitives" such as Poseidon hash, MiMC sponge hash, Pedersen commitment, or Kate commitment, the efficiency can be greatly increased. For Pedersen commitment, it's almost 6 times faster in Gabizon19 than Groth16. For MiMC hash, it's still 3 times faster. As shown in Figure 3, Gabizon's construction is also better scalable according to the data from Aztec [59].

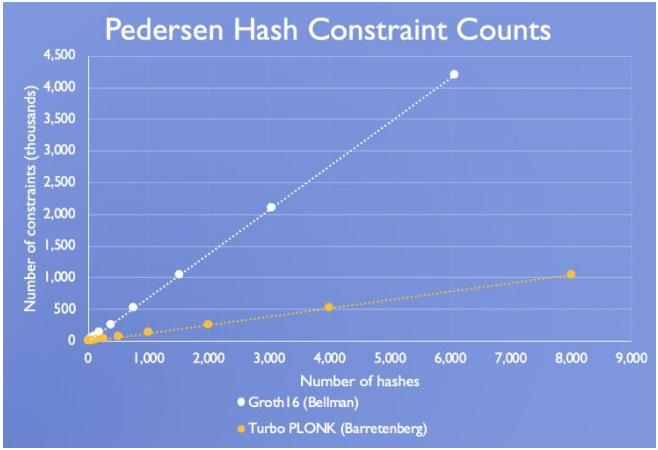

Fig. 3. Circuit constraints versus the number of hashes [59].

### B. System Performance

We test our setup using the following construction:

- VMware 15 Ubuntu 20.04 LTS 64bit 2 vCore 8GB RAM, under AMD 2700x physical machine.
- Ubuntu Server 20.10 64bit, under Raspberry Pi 3 model BCM2837 with 1GB RAM.

The performance benchmark was performed with a rust implementation of Gabizon19's work [60]. The benchmark runs on Poseidon hash. For a $2^{20}$ Merkle tree, the Poseidon hash has about 4,380, or ~$2^{12}$ constraints in the circuit for Merkle path verification[61]. Table III shows the performance on various system settings.

**Table III: Comparison of Performance on Physical Machine**

|  | Device | Proof Generation | Verification | Proof Size |
|---|---|---|---|---|
| Gabay et al. [27] | Raspberry Pi 3 | 14,380 ms | trivial | 128 bytes |
| Huang et al. [30] | Samsung Galaxy S4 | 2,000 ms | trivial | 1205 bytes |
| Ho and Lin | Raspberry Pi 3 | 1,231 ms *1 | trivial | 1208 bytes |
| Ho and Lin | AMD2700x | 230 ms *1 | trivial | 1208 bytes |

[1] Proportional to Merkle tree depth.

Gabay's work combined Ethereum blockchain onto the system to achieve data integrity and mutual trust between EV charging providers and EV users. However, the property is not needed in our system, as one does not care (and shouldn't be able to) explore the raw authentication data on the identity commitment tree maintained by the authorities – he / she only cares if their authentication request is rendered valid or not. Gabay's system also designed an *administrative action*, where a privileged admin can interfere and reveal user's identity when a dispute happens. This is clearly not ideal from the standpoint of user privacy.

Huang's work designed a novel zk-PoK system based on Lee et al.'s work [62]. The system assumed an in-door closed parking space where its parking lot terminal and parking service provider are rather powerful, i5-7200U, and i7-6700K, respectively. In an outdoor environment such as on-street parking in our scenario, it's insufficient to assume rather powerful computational resources for economical purposes. The zk-PoK system, while providing more security features than our construction, is also more computationally intensive.

Our construction is rather minimal compared to others, as we aimed to achieve different properties with our system. We designed a one-way authentication with *zero-knowledge proof* as our main feature, which prevents users from being associated between sessions. It works similarly as a one-time password, but with untraceability.

## VI. DISCUSSIONS

To recap, our system applied Gabizon's zk-SNARK variant, combined with the commitment scheme and Merkle tree to achieve zero-knowledge set membership proof. The privacy of the user is thus preserved as his / her login between different sessions cannot be linked thanks to the zero-knowledge provided by the protocol.

However, there's still some works that can be done in the future, namely:

- The security of Gabizion's work could be challenged in the future, such as malleability[63], as the work's still relatively new.

Groth's work in 2016 has such an issue. However, it is still not clear that if Gabizons' work suffers the same vulnerability.

- The system only considered one-way authentication, which authenticates user's membership proof from the server.

The mutual authentication is not considered in our scheme, as it does not harm user in any way even if the authentication request was intercepted by an adversary – the proof is rendered invalid after the next nullifier period, and the adversary cannot forge another proof and convince the server to accept it. The design philosophy also makes the zk proof in a way, worked as a one-time password, which ensures the complexity and the security of the zk authentication scheme suits our need in the on-street parking scenario.

- The server could record user's Merkle tree query request, and link the following authentication request with the former query request, as it possesses a high probability that the two communications are from the same user.

Such side-channel attack is hard to avoid in our system. To prevent such an issue, the user can run a "full node" which keeps syncing the Merkle tree data and the rotating nullifier from the server, which prevents the upcoming authentication request from associated with the query request. Such solution implies high storage and network cost on the user-side, and is not recommended.

- The use of physical camera is not covered in the scope of our system, as it contradicts the design rationale.

The design rationale of the system: raise the cost to citizen profiling, while still able to authenticate and manage on-street parking usage. Currently, it's relatively costly to mass-capture, record, and classifies car license plates. However, it's relatively easy to enter a single line of SQL command in the database to rebuild certain user's usage traces. This is precisely the reason why we raised the hurdle of the latter one by anonymized user traces between sessions.

- User accountability is seemingly not considered, which is not the case.

Some might argue that it's impractical to anonymize user, for the case that sometimes it's necessary to chase certain user for his improper behavior during the use of the system. However, most of the improper behaviors are conducted physically by users, and the authority always needs to intervene with human labor. That is, it's a non-factor that whether the system recorded user's access log or not, would never affect the accountability of the system – the authority always has to dispatch people to the site physically whenever undesirable demeanor happens, thus revealing the license plate and the identity of the user.

- The parking metering is intentionally excluded in our system

At the moment [64]–[66], it's not possible to hide, unlink user's identity from different sessions, as it does not satisfy the basic requirement of metering – accumulate certain user's usage statistics. The cited work, however, obfuscated the meter readings, which in return protects user's privacy to an extent. Balasch et al. [66] applied a payment channel called optimistic payment in his system, to hide the location data and prices data while still able to compute the correct fees.

We suggest a simple payment system to be implemented in our system, albeit primitive, to inspire people onto further exploration of such possibilities. Consider each identity commitment registered as a one-time redeemable ticket. While registering to the authority, user must send together with a payment proof in order to prove that he / she has already paid a fixed amount of fee for a 1-hour-long parking. User can send multiple identity commitments with the same identity (*UID*), to gain the permission to redeem such parking tickets later. When parking, user can send such proof corresponding to the identity commitment (redeemable ticket), and the server will nullify the one-time redeemable ticket and accept a 1-hour-parking from the user. The process shall be repeated by the user or the OBU as the eligible parking time expired.

For other usages of *zero-knowledge membership proof* in the smart city, Yeh [67] provided some practical scenarios such as whistle-blowing, anonymous goods distribution, or public identity infrastructure. Such explorations are yet to be implemented, and are worth pursuing for the sake of privacy protection.

## VII. CONCLUSION

We presented an anonymous on-street parking authentication scheme via zero-knowledge set membership proof, improving the vulnerability of identity forgery, replay attack and masquerade attack from our previous results. For the on-street parking scheme, most of the current research does not consider privacy and identity traces leaking server-side. That is, one can either social engineering employees from the authority, or conduct an APT attack – infiltrate the system, and patiently gather user traces to profile them from inside.

To protect the system from both inside and outside adversaries, our approach made user's identity unlinkable between login sessions, while still be able to authorize the user properly, thanks to the zero-knowledge set membership proof combined with Merkle tree and the Plonk proof system. This further protects user privacy from being harmed in the mass surveillance era, as one cannot easily gather user traces by a single line of SQL command anymore in our system, compared to other currently existing solutions.

We'd like to raise awareness for the privacy topic, urging researchers and engineers to implement inherently better security and privacy solutions to protect citizens in the landscape of the smart city. Such solutions should be "fool-proof" – in the aviation industry, almost all the accidents led to improvements about the **systematic failure**, rather than blaming on individuals. In our case, a zero-knowledge set membership proof based authentication system prevents human misuse of the database access permission, and security breaches from outside adversaries. User's privacy is then well-protected even after the security incident happened – this should be THE target we pursue for smart city applications.

## ACKNOWLEDGMENT

The framework structure and protocol design of this paper was inspired by the following researches: "Zerocash: Decentralized Anonymous Payments from Bitcoin", and "Community Proposal:Semaphore: Zero-Knowledge Signaling on Ethereum".

The authors of this paper would also like to thank Ting Yi Huang for her contribution to the drawing of charts in this paper.